# Specification Portability Across LLM Development Agents: Cross-Agent Compatibility in Specification-Driven Software Migration

Oleg Grynets
EPAM Systems
McLean, Virginia, USA
oleg_grynets@epam.com

Oleksii Ilchuk
EPAM Systems
Lviv, Ukraine
oleksii_ilchuk@epam.com

Dariia Zatulna
EPAM Systems
Lviv, Ukraine
dariia_zatulna@epam.com

Vasyl Lyashkevych
EPAM Systems
Lviv, Ukraine
vasyl_lyashkevych@epam.com

***Abstract*—Large Language Models (LLMs) are increasingly used in specification-driven software development, where specifications become operational artifacts that directly guide implementation agents. This raises an important interoperability question: whether a specification created or structured within one LLM development environment can be reliably consumed by another agent without degrading implementation quality. This paper investigates cross-agent specification portability using Oracle-to-PostgreSQL migration as a controlled software transformation task. The study combines two experimental stages. First, a specification-first migration pipeline was evaluated on 1,006 PL/SQL files, of which 623 were successfully regenerated and 380 generated scripts executed successfully in PostgreSQL 16. Second, cross-agent experiments were conducted on a dataset of 1,802 Oracle scripts with corresponding PostgreSQL implementations using Amazon Kiro, Google Gemini, and GitHub Copilot, with Claude Code and Cursor included in the initial single-agent evaluation. Native and foreign specifications were assessed using Token F1, exact match, SQL syntax validity, AST exact match, AST mean similarity, and immediate runnability. The results show that specification size alone does not predict implementation quality and that cross-agent transfer can produce substantial agent-dependent degradation. The strongest replicated case occurred when Gemini directly consumed a Kiro-origin specification, producing a Token F1 of 0.035, SQL syntax validity of 2.33%, and AST mean similarity of 0.015. Rewrite, compression, and retrieval-augmented specification ingestion were further evaluated as adaptation strategies. Rewriting substantially improved Gemini in the tested configuration, compression did not provide a universal benefit, and retrieval-augmented ingestion was the only common strategy represented on the per-agent Pareto frontiers of both Gemini and Copilot. The findings suggest that specifications in heterogeneous SDD workflows should not automatically be treated as agent-neutral artifacts and motivate explicit consideration of specification portability, agent-specific interpretation, and retrieval-based access in multi-agent software engineering.**

## I. Introduction

Large language models (LLMs) are increasingly integrated into software engineering workflows for code generation, translation, program repair, documentation, testing, and software modernization. This transition is accompanied by a shift from conventional prompt-based assistance toward agentic development, in which an LLM-based agent can interpret a high-level task, decompose it, invoke tools, generate artifacts, and iteratively refine an implementation [1]. Specification-driven development (SDD) extends this transition by making a specification an operational input to the development process rather than only a descriptive document. The idea has historical roots in Agile Specification-Driven Development, where specifications, contracts, and tests were integrated into the development workflow [2]. In contemporary LLM-assisted development, however, specifications may directly guide autonomous code generation and therefore become an intermediate representation between software intent and implementation [3].

This change is particularly relevant to software migration. Database migration between heterogeneous database management systems requires preservation of application semantics while transforming dialect-specific constructs, data types, built-in functions, procedural logic, triggers, exception handling, sequences, packages, and other platform-dependent elements. LLM-supported SQL translation can automate portions of this work, but direct source-to-target generation may introduce semantic drift and reduce transformation traceability [4]. A specification-based Code–Text–Code approach addresses this limitation by introducing an intermediate textual representation that captures behavior, identifiers, computational flow, conditions, dependencies, side effects, and domain intent before target code is generated [5]. In this interpretation, migration becomes a controlled transformation through an intermediate specification rather than direct syntactic translation.

The importance of the intermediate representation is also supported by related work on architecture-level knowledge representation and specification-controlled software generation. A unified architecture metamodel can provide structured context across representation layers [6], while security-oriented SDD research shows that information required for downstream implementation must be made operational and traceable in a specification rather than remaining implicit [7]. Token-optimization experiments in Oracle-to-PostgreSQL migration further demonstrate that reducing context is not equivalent to preserving useful context: aggressive information reduction may increase efficiency while damaging semantic preservation, whereas selective strategies may retain substantially more transformation quality [8]. These findings motivate investigation of not only what information is contained in a specification, but also how that information is represented and consumed.

The problem becomes more complex in multi-agent software engineering. An implementation workflow may use different agents for requirements processing, specification generation, implementation, testing, migration, or validation. Dynamic orchestration research consequently treats agents as specialized components whose roles and interaction paths may change at runtime [9]. Specification-driven synthesis research similarly treats the specification as a first-class artifact from which generated software artifacts can be produced and regenerated [10]. At the same time, studies of LLM-mediated software development identify security, reliability, and behavioral risks associated with generated code [11], reinforcing the need for controlled intermediate representations and validation. Thus, heterogeneous agent workflows introduce an additional question that is not addressed by merely improving a single model or a single prompt: whether a specification prepared in one agent environment remains operationally effective when transferred to another.

This question is important because current empirical evidence on LLM-based software engineering consistently shows strong sensitivity to context and task structure. A systematic review of LLM applications in software engineering identifies code generation as one of the dominant research areas while also emphasizing unresolved challenges related to reliability, evaluation, context, and practical software-engineering integration [13]. Empirical research on LLM code generation additionally demonstrates substantial nondeterminism: repeated requests can produce structurally and behaviorally different implementations even for the same task, which complicates reproducibility and the interpretation of single-run comparisons [14]. These findings imply that variation observed after transferring a specification cannot automatically be attributed to specification quality alone.

The difficulty grows as the generated artifact becomes larger and more dependent on external context. SWE-bench demonstrated that real-world software engineering tasks require models to coordinate changes across functions, classes, files, execution environments, and long contexts [15]. EvalPlus similarly showed that apparently correct generated programs may fail when subjected to substantially expanded functional testing [16]. At class level, ClassEval found that model performance decreases when generation moves from isolated methods to collections of interdependent program elements [17]. DevEval extends this observation to real repositories, where requirements and implementation depend on repository-level structure and dependencies [18]. These studies collectively indicate that lexical similarity or isolated code-generation success is insufficient for evaluating complex LLM-mediated transformations.

Context availability is another critical factor. CrossCodeEval demonstrates that cross-file code completion becomes substantially more difficult when relevant external context is absent and that providing appropriate context improves completion quality [19]. RepoCoder addresses this problem using iterative retrieval and generation at repository level [20], while Repoformer shows that retrieval itself should be selective because irrelevant retrieved context can reduce robustness and efficiency [21]. These results are directly relevant to specification ingestion: a long specification should not necessarily be interpreted as a single undifferentiated context, and selective access to its relevant portions may constitute an alternative to complete direct ingestion.

The emergence of multi-agent development frameworks makes specification transfer even more consequential. ChatDev distributes software-development activities among communicating agents participating in design, coding, and testing [22]. MetaGPT similarly organizes specialized agents around standardized operating procedures and intermediate artifacts [23]. Such architectures demonstrate that software generation no longer necessarily occurs inside a single agent/model interaction. Instead, artifacts created by one role can become inputs to another. This makes the interoperability of intermediate artifacts—requirements, specifications, designs, tests, and code—a practical multi-agent software-engineering problem.

Requirements representation also influences generated behavior. Test-driven interactive code generation has been proposed as a mechanism for formalizing ambiguous intent before accepting generated implementations [24]. Repository-level code question answering studies further show that retrieval and self-alignment can improve an LLM's access to distributed software context [25]. Retrieval-augmented code generation has also demonstrated benefits for programming languages with limited training resources when appropriate examples or documentation are retrieved [26]. ClarifyGPT, in turn, showed that ambiguous natural-language requirements can lead to divergent implementations and that explicit clarification of the requirement can improve generated-code correctness [27]. Collectively, these studies indicate that generation quality depends not only on model capability but also on the structure, completeness, accessibility, and interpretation of the information provided to the model.

Despite this progress, existing research primarily evaluates model capability, repository context, retrieval mechanisms, requirements clarification, or collaboration among agents. The portability of a complete software specification between heterogeneous LLM development agents remains insufficiently characterized. In particular, it is unclear whether a specification that produces acceptable implementation behavior with the agent for which it was generated will retain this effectiveness when consumed by another implementation agent, and whether adaptation strategies can reduce incompatibility.

The present study investigates this problem using Oracle-to-PostgreSQL migration as a controlled software-transformation task. The experimental study consists of two complementary stages. In the first study, a specification-first Oracle-to-PostgreSQL pipeline was evaluated on 1,006 PL/SQL source files. Of these, 623 files were successfully regenerated, and 380 of the regenerated files subsequently executed successfully in a PostgreSQL 16 environment. The experiment also showed substantial differences among database-object categories, with tables reaching approximately 85% regeneration success while procedural and dependency-sensitive constructs remained more difficult.

The second study extended the experiment to cross-agent specification transfer. A dataset of 1,802 Oracle scripts with corresponding PostgreSQL scripts was used to evaluate specifications and implementations involving Amazon Kiro, Google Gemini, GitHub Copilot, and, during the initial stage, Claude Code and Cursor. The measured dimensions included Token F1, exact match, SQL syntax validity, AST exact match, AST mean similarity, and immediate runnability. Native and foreign specifications were compared, followed by experiments with specification rewriting, summarization/compression, and retrieval-augmented ingestion. The resulting measurements showed that specification length alone did not explain code-generation quality and that the effect of using a specification originating from another agent was strongly agent-dependent. In particular, the substantial degradation observed when Gemini directly consumed a Kiro-origin specification was reproduced in a second run. Retrieval-augmented ingestion did not dominate every individual metric but was the only common ingestion strategy represented on the per-agent Pareto frontiers reported for both Gemini and Copilot.

Based on these observations, this paper addresses the following research questions:

RQ1: How does the origin of a specification affect implementation quality when the specification is transferred between heterogeneous LLM development agents?

RQ2: Are differences in generated-code quality sufficiently explained by specification size, or do specification representation and agent-specific interpretation also affect the result?

RQ3: To what extent can rewriting, compression, or retrieval-augmented ingestion mitigate degradation associated with a foreign specification?

RQ4: Which specification-ingestion strategy provides the most robust trade-off across the evaluated implementation agents?

The contribution of the study is therefore empirical rather than a claim of universal agent incompatibility. The paper (i) combines a specification-driven Oracle-to-PostgreSQL migration pipeline with controlled cross-agent specification-transfer experiments; (ii) evaluates native and foreign specifications using lexical, syntactic, structural, and operational metrics; (iii) compares direct, rewritten, compressed, and retrieval-augmented specification ingestion using only the measured experimental configurations; and (iv) identifies specification portability as an open problem for heterogeneous specification-driven development.

## II. Related Work

### A. *Specification-Driven and Specification-Based Software Development*

Specification-driven development has its foundations in software-engineering approaches in which explicit specifications constrain implementation and verification. Ostroff et al. combined test-driven development and design by contract within Agile Specification-Driven Development [2]. The recent adoption of LLM agents substantially changes the role of these specifications because they can now be consumed directly during code generation rather than serving exclusively as human-oriented artifacts [3].

For migration and reengineering, an intermediate specification can also separate source-language representation from target-language generation. The Specification-Based Code–Text–Code approach transforms source code into a neutral textual description before generating the target implementation [5]. The specification can capture program behavior, identifiers, control conditions, side effects, data dependencies, and domain-specific intent, thereby reducing reliance on surface syntax. This concept is particularly relevant to Oracle-to-PostgreSQL migration because source and target systems share SQL foundations while differing substantially in dialect-specific and procedural constructs.

Architecture-oriented work further indicates that generated software artifacts can benefit from structured representations at several abstraction levels [6]. Similarly, specification-driven intelligent-monitoring synthesis treats the specification as an explicit artifact from which implementation elements are generated and regenerated [10]. Security Knowledge Transition research extends this perspective by representing relations between software entities, threats, risks, requirements, controls, implementation rules, and verification evidence explicitly in an SDD artifact [7]. These studies collectively motivate treating the specification as an operational knowledge representation rather than merely a large natural-language prompt.

However, none of these works establishes that a specification representation remains equally effective across heterogeneous implementation agents. The present study therefore focuses on the transition: “Source artifact → specification → heterogeneous implementation agent → target artifact”.

The empirical question is not whether specifications are useful in principle, but whether their effectiveness is preserved when the producer and consumer environments differ.

### B. *LLM Code Generation and Evaluation*

Research on LLMs for software engineering has expanded rapidly. The systematic review by Hou et al. organizes existing LLM applications across software-engineering tasks and identifies code generation, testing, maintenance, and program understanding as major research directions while emphasizing open issues related to robustness and evaluation [13].

A major limitation in LLM-based code-generation experiments is nondeterministic behavior. Ouyang et al. evaluated 829 code-generation problems across CodeContests, APPS, and HumanEval and found considerable output variation across repeated requests; importantly, setting temperature to zero did not eliminate nondeterminism [14]. This is methodologically relevant to cross-agent comparison and supports the need to interpret individual cross-agent runs cautiously and to replicate anomalous results.

Evaluation methodology itself is also important. Liu et al. introduced EvalPlus to extend the test coverage of established program-synthesis benchmarks and demonstrated that generated programs considered correct under limited tests may fail additional functional cases [16]. Similarly, Du et al. introduced ClassEval for class-level generation and showed that models perform substantially worse when tasks include several interdependent code units rather than isolated functions [17].

Repository-oriented benchmarks strengthen this observation. SWE-bench contains 2,294 real software-engineering problems derived from GitHub issues and requires models to reason over repositories and execution environments [15]. DevEval provides 1,825 samples from 115 repositories with requirements, source repositories, reference code, and dependency annotations and demonstrates the additional difficulty of generation in realistic repository contexts [18]. These works support the multidimensional evaluation adopted in the present study: Token F1 and exact match alone cannot establish that migrated code is syntactically, structurally, or operationally correct.

Accordingly, the provided migration experiments use several complementary measures: Token F1, exact match, SQL syntax validity, AST exact match, AST mean similarity, and immediate runnability. These metrics are retained exactly as defined in the supplied experimental study and are not supplemented here with unmeasured performance results.

### *C. Context, Retrieval, and Specification Ingestion*

A central issue in repository-level generation is identifying which information should be exposed to the model. CrossCodeEval shows that generation quality is strongly affected by access to cross-file dependencies and provides a benchmark explicitly requiring such external context [19]. RepoCoder combines similarity-based retrieval with iterative retrieval and generation and reports improvements over in-file and simpler retrieval baselines [20].

Repoformer adds an important qualification: retrieval is not always beneficial. Its selective-retrieval architecture is motivated by evidence that irrelevant or noisy retrieved context can harm code completion, and it learns whether retrieval should be applied for a particular generation step [21]. This observation is particularly relevant to long specifications. More available specification text does not necessarily imply more useful context for a particular implementation operation.

Repository-level code QA also benefits from retrieval-aware techniques. Strich et al. combine self-alignment and RAG pipelines to improve understanding of source semantics, file dependencies, and repository-level metadata [25]. Dutta et al. similarly investigate retrieval-augmented code generation in low-resource programming languages, where relevant examples and documentation provide information unavailable from model parameters alone [26].

These works provide literature support for evaluating retrieval in the present study, while the performance claims for RAG are based on the cross-agent experiment reported in this paper. In that experiment, agents were given retrieval tools for specification access and prevented from consuming the complete specification directly. The measured results did not establish RAG as universally superior on every metric; rather, RAG was the only common evaluated strategy that appeared on the reported per-agent Pareto frontiers for both Gemini and Copilot.

This distinction is important: repository-level RAG literature motivates the strategy, whereas the Oracle-to-PostgreSQL experiments reported here determine its observed effectiveness.

### *D. Multi-Agent Software Engineering*

Modern LLM-based software engineering increasingly distributes work among specialized agents. ChatDev models software development through communicating agents participating in design, coding, and testing and treats agent communication as a mechanism for coordinating specialized development roles [22]. MetaGPT similarly uses specialized agents and standardized operating procedures to organize collaborative problem solving and software production [23].

Related orchestration work explores dynamic composition of specialized agents, including runtime decomposition and the recruitment of new agents when a fixed workflow is insufficient [9]. These systems establish the broader architectural motivation for the present research: in an agentic software-development pipeline, the agent generating an intermediate artifact does not necessarily have to be the same agent consuming it.

Nevertheless, multi-agent studies generally concentrate on task decomposition, communication protocols, role specialization, workflow organization, and collective quality. They do not directly quantify whether a specification generated under one development agent/framework remains an equally effective implementation artifact for another agent. The cross-agent experiments in this study address this narrower issue by explicitly exchanging specifications between implementation environments.

### *E. Requirements Ambiguity, Intent, and Intermediate Representations*

The effectiveness of generated code depends on the quality and determinacy of the information provided to a model. Fakhoury et al. proposed test-driven interactive code generation in which tests help distinguish alternative behaviors that may arise from ambiguous intent [24]. ClarifyGPT likewise detects ambiguous requirements and generates clarification questions before final code generation; its empirical evaluation showed that requirement clarification can improve code-generation performance [27].

These studies are relevant because specifications are intended to reduce the ambiguity of direct natural-language requests. However, eliminating semantic ambiguity at the requirements level does not automatically establish representation invariance across agents. Two specifications may describe the same target functionality while organizing instructions, constraints, dependencies, examples, and implementation guidance differently.

This issue is visible in the supplied cross-agent experiments. Kiro generated the largest specification, approximately 1,597 lines, whereas Gemini generated a considerably smaller specification, approximately 193 lines. Nevertheless, larger specification size did not correspond to universally stronger implementation results. Furthermore, asking Gemini to rewrite a Kiro-origin specification into its preferred representation substantially changed the measured outcome, whereas the same strategy did not provide a universal improvement for Copilot. These results motivate separating specification completeness from agent-specific specification interpretability.

### *F. Database Migration and Context Optimization*

Database dialect migration provides an especially useful case for studying specification portability because transformation correctness includes both syntactic and semantic constraints. Direct SQL translation can be complicated by dialect-specific constructs and by the need to preserve procedural behavior [4]. The Specification-Based Code–Text–Code approach motivates an explicit intermediate representation for such transformation [5].

Existing token-optimization experiments for Oracle-to-PostgreSQL migration further show that context manipulation should be treated as a quality-cost trade-off rather than simple shortening [8]. That study evaluated several transformations and showed that aggressive reduction can remove semantically important information. The present cross-agent experiment addresses a different but related problem. Instead of optimizing token count as the primary objective, it compares several ways of exposing a foreign specification to an implementation agent: direct ingestion, rewriting, summarization/compression, and retrieval-augmented access.

### G. Research Gap

The literature establishes five relevant observations.

First, specifications and intermediate representations can provide additional control and traceability in LLM-mediated software transformation [2], [5]–[7]. Second, code-generation quality decreases as tasks require broader structural and repository context [15], [17]–[19]. Third, retrieval can improve access to relevant distributed context, but indiscriminate retrieval can also introduce ineffective or harmful information [20], [21], [25], [26]. Fourth, multi-agent frameworks increasingly pass development artifacts among specialized agents [22], [23]. Fifth, ambiguity and representation of requirements materially affect generated-code behavior [24], [27].

However, these lines of research do not establish whether an SDD specification can be considered agent-neutral.

The remaining gap can therefore be formulated as follows: *"Existing LLM-based software-engineering research studies specification quality, code-generation capability, context retrieval, requirements clarification, repository-level generation, and multi-agent collaboration, but provides limited empirical evidence on whether a specification originating from one LLM development-agent environment preserves its implementation effectiveness when consumed by another heterogeneous agent"*.

The present study addresses this gap through native- and foreign-specification experiments in an Oracle-to-PostgreSQL migration setting. Its conclusions are deliberately restricted to the observed agents, specifications, ingestion strategies, dataset, and metrics. The study does not claim that foreign specifications universally decrease generation quality; rather, it evaluates whether cross-agent specification effects exist in the supplied experimental setting and which of the tested ingestion strategies are most robust under those conditions.

## III. Research Problem

Current SDD implicitly encourages the conceptual relation $S \rightarrow I$, where specification $S$ determines implementation $I$.

For heterogeneous agentic development, however, we propose the extended relationship:

$$I = G(S, A, I_s, C), \tag{1}$$

where $S$ – specification; $A$ – implementation agent; $I_s$ – specification-ingestion strategy; $C$ – task and contextual conditions; $G$ – implementation-generation process.

When the specification itself is generated by another agent:

$$S = S(A_g, R, K) \tag{2}$$

where $A_g$ – specification-generating agent; $R$ – source requirements or source artifact; $K$ – available contextual knowledge.

Thus:

$$Q_{code} = f(A_g, A_i, S, I_s, C) \tag{3}$$

where $Q_{code}$ is the resulting implementation quality.

The central hypothesis is therefore:

H1: Specifications generated within different LLM agent environments are not universally portable across implementation agents.

Three secondary hypotheses are considered:

H2: Specification length alone is not a sufficient predictor of generated implementation quality.

H3: Representation adaptation can reduce cross-agent compatibility loss, but its effectiveness is agent-dependent.

H4: Retrieval-based specification ingestion provides a comparatively robust strategy across the evaluated heterogeneous agents.

## IV. Cross-Agent Specification Portability

We define specification portability as the capability of a specification produced or organized under one development-agent environment to preserve its implementation effectiveness when consumed by another development agent.

Let $Q_m(A_j, S_i)$ represent implementation quality obtained by agent $A_j$ using a specification originating from agent $A_i$. The native quality is:

$$Q_m(A_j, S_j). \tag{4}$$

Because implementation quality is evaluated through multiple metrics, cross-agent compatibility is defined separately for each quality metric m:

$$C^{(m)}_{i \rightarrow j} = \frac{Q_m(A_j, S_i)}{Q_m(A_j, S_j) + \epsilon}, \tag{5}$$

where $Q_m$ denotes the value of quality metric $m$, such as Token F1, SQL Syntax Validity, AST Mean Similarity, or Immediate Runnability; $\epsilon$ prevents numerical instability for metrics approaching zero.

Interpretation is straightforward:

$$C^{(m)}_{i \rightarrow j} \approx 1, \tag{6}$$

indicates approximate portability:

$$C^{(m)}_{i \rightarrow j} < 1, \tag{7}$$

indicates compatibility loss; and

$$C^{(m)}_{i \rightarrow j} > 1, \tag{8}$$

indicates that the foreign representation provides an advantage for the target agent.

Because generation quality is multidimensional, $Q$ should not normally be represented by a single raw metric. We define:

$$Q = \langle F_1, E, V_{SQL}, M_{AST}, S_{AST}, R \rangle, \quad (9)$$

where $F_1$ – Token $F_1$; $E$ – exact-match rate; $V_{SQL}$ – SQL syntax validity; $M_{AST}$ – AST exact-match rate; $S_{AST}$ – AST similarity; $R$ – immediate runnability.

This yields a compatibility matrix:

$$\mathbf{C} = \begin{bmatrix} C_{1\to1} & C_{1\to2} & \cdots \\ C_{2\to1} & C_{2\to2} & \cdots \\ \vdots & \vdots & \ddots \end{bmatrix}. \quad (10)$$

Importantly:

$$C_{i\to j} \neq C_{j\to i}. \quad (11)$$

in general.

Cross-agent compatibility should therefore be regarded as directional, not symmetric.

This interpretation corresponds closely to the empirical observations obtained in the experiments.

## V. Methodology

### A. Stage 1: Specification-Driven Migration Pipeline

The overall specification-driven migration workflow is illustrated in Fig. 1. Source SQL is first preprocessed and decomposed into database objects before the agent generates an intermediate specification and subsequently derives the target PostgreSQL implementation.

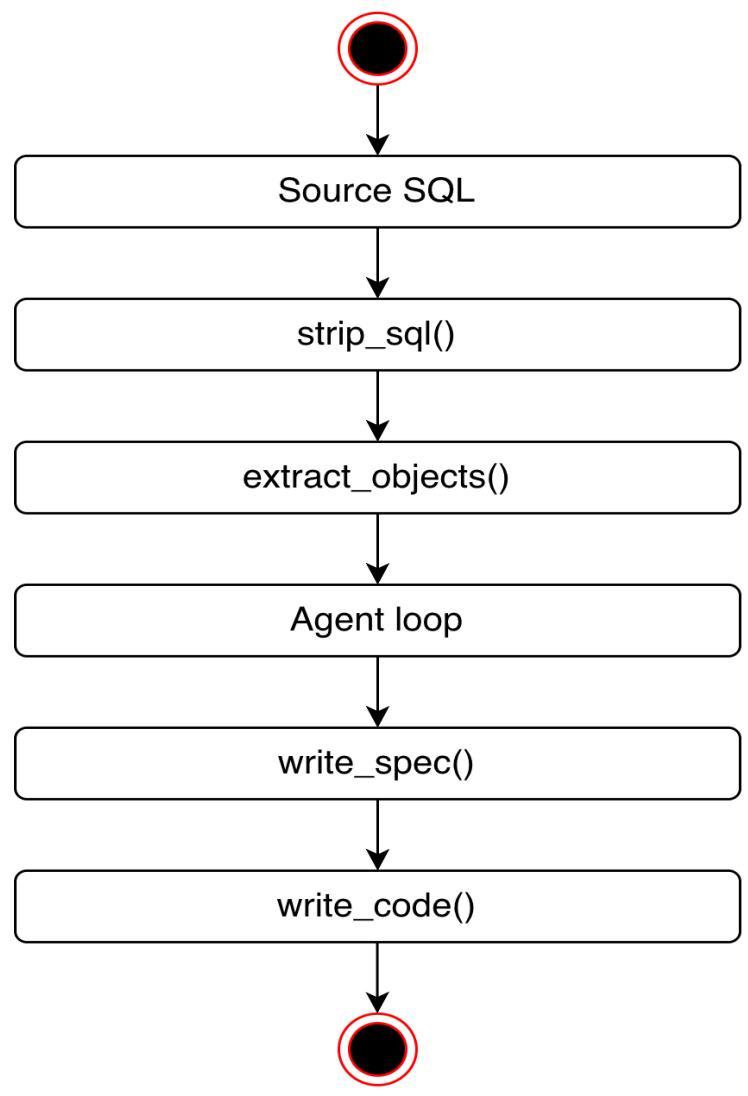

Fig. 1. Specification-driven Oracle-to-PostgreSQL migration pipeline.

As shown in Fig. 1, preprocessing separates deterministic source-code preparation from LLM-mediated specification and code generation. This object-level decomposition reduces unnecessary input context and allows individual migration objects to be regenerated independently when conversion fails.

Object-level decomposition has two purposes.

First, irrelevant input is removed, reducing token consumption.

Second, each independently generated object receives its own specification. Consequently, if a generated target object is incorrect, only that object must be regenerated instead of the complete source file.

The agent loop then produces a specification and derives PostgreSQL code from that specification. The internal execution of the agent loop is shown in Fig. 2. The sequence coordinates the application layer, agent controller, LLM, and MCP-based tools while controlling iterative tool invocation and termination.

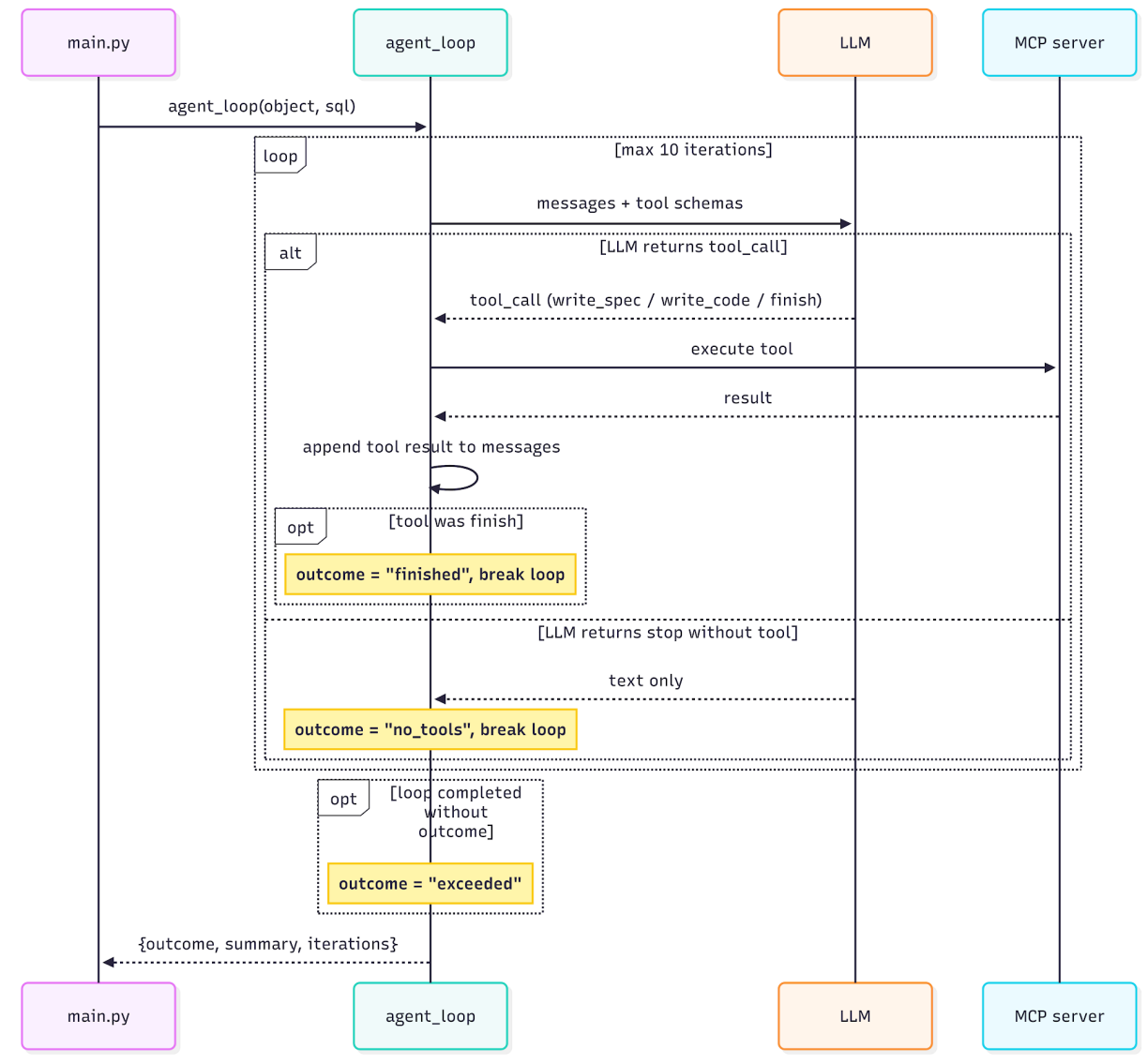

Fig. 2. Sequence of interactions within the specification-driven agent loop.

The loop repeatedly supplies the model with messages and available tool schemas, executes requested tools, and returns their results to the agent context. Execution terminates when the agent explicitly invokes the finishing operation, returns no further tool calls, or reaches the iteration limit. This mechanism supports iterative specification and code refinement while maintaining explicit execution control.

Two models were evaluated in the initial implementation:

GPT-oss-120b;

GPT-o4-mini.

Average token consumption was approximately 2,000 tokens per iteration, with approximately three iterations on average; smaller objects could require approximately 900 tokens and larger ones up to approximately 3,000.

### B. Stage 2: Cross-Agent Experiment

The second stage changed the experimental unit from migration objects to the relationship between specification origin and implementation agent.

Before cross-agent transfer was introduced, each development environment was evaluated in a native single-agent workflow in which specification generation, human review, code generation, and testing were performed iteratively within the same agent environment. This baseline workflow is shown in Fig. 3.

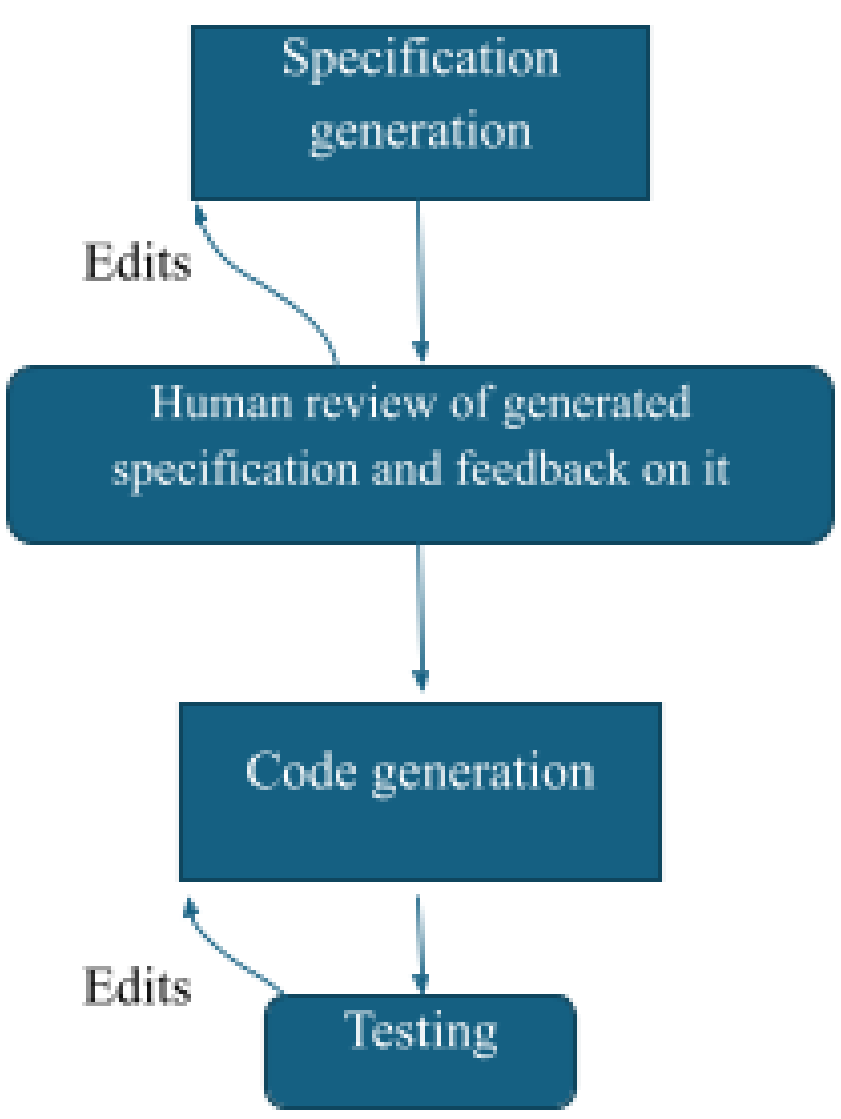

Fig. 3. Native single-agent specification-driven development workflow used in the preliminary evaluation.

The workflow establishes the native condition in which the same agent environment is responsible for producing and subsequently consuming the specification.

Because the original single-agent experiment was not initially designed as a controlled baseline for cross-agent comparison, the native configurations of Kiro, Gemini, and Copilot were rerun before the cross-agent experiments. The resulting baseline measurements are summarized in Table I.

TABLE I. NATIVE-SPECIFICATION BASELINE PERFORMANCE

| Framework + Agent | Token F1 | Exact Match, % | SQL Syntax Validity, % | AST Exact Match, % | AST Mean Similarity | Runnable, % |
|---|---|---|---|---|---|---|
| Amazon Kiro | 0.3652 | 0.33 | 33.74 | 0.33 | 0.2073 | 0.72 |
| Spec Kit + GitHub Copilot | 0.75 | 13.15 | 43.51 | 12.76 | 0.2566 | 0 |
| Antigravity + Gemini | 0.73 | 13.32 | 45.54 | 12.98 | 0.69 | 2.22 |

The later cross-agent experiments modify this condition by retaining the specification artifact while changing the implementation agent.

Five environments were initially investigated:

- Amazon Kiro;
- Google Gemini / Antigravity;
- GitHub Copilot / Spec Kit;
- Claude Code;
- Cursor.

Because of pricing constraints, the systematic cross-agent stage concentrated primarily on Kiro, Gemini, and Copilot.

Agents received the same functional request for a deterministic Oracle-to-PostgreSQL migration utility.

The requested system included support for:

- SQL datatype conversions;
- built-in functions;
- sequences;
- set operators;
- stored procedures;
- functions;
- triggers;
- packages;
- cursors;
- collections;
- exception handlers;
- preservation of unsupported Oracle constructs;
- remediation notes;
- structured conversion reports;
- optional reference-database access;
- CLI operation;
- AST/grammar-based structural parsing;
- unit and property-based testing.

The evaluation dataset contained 1,802 Oracle scripts with corresponding PostgreSQL scripts, varying in length and complexity.

## VI. EVALUATION METRICS

The following metrics were used.

**1. Token F1**. Token-level similarity between generated and expected outputs:

$$F1 = \frac{2PR}{P+R}. \qquad (12)$$

**2. Exact Match.** Percentage of generated outputs identical to expected target scripts.

**3. SQL Syntax Validity.** Percentage of outputs recognized as syntactically valid SQL using sqlglot.

**4. Immediate Runnability.** Percentage of generated scripts that could execute without additional manual modifications in the experimental database environment.

The original study explicitly notes that this value should be interpreted as a lower bound, because testing was performed against an empty live database with generated stubs rather than the exact production schema.

**5. AST Exact Match.** Percentage of converted scripts whose parsed abstract syntax tree exactly matched the expected output.

**6. AST Mean Similarity.** Average structural similarity between generated and expected ASTs.

These measures intentionally distinguish lexical, syntactic, structural, and operational quality.

## VII. RESULTS

### *A. Feasibility of Specification-Driven Migration*

The initial specification-driven pipeline processed:

$$N = 1006$$

PL/SQL source files.

Of these,

$$623/1006 \approx 61.9\%,$$

were successfully regenerated.

Execution in PostgreSQL 16 subsequently showed that:

$$380/623 \approx 61.0\%$$

of generated scripts executed successfully.

Tables were the most successful object category, with an approximately 85% regeneration success rate, closely followed by views and triggers. Queries performed worst in that experiment, and procedures, functions, indexes, and other schema-dependent constructs remained substantially more difficult.

These results established that SDD-based migration was feasible but did not yet explain what properties of a specification contributed to successful downstream generation.

### *B. Specification Size Does Not Explain Quality*

The first cross-agent observation concerns specification size. Kiro produced the largest specification, approximately 1,597 lines, whereas Gemini produced a specification of approximately 193 lines. Nevertheless, the largest specification did not lead to the strongest implementation across evaluated metrics.

After tuning, the reported results are shown in Table II.

The result provides direct evidence against a simplistic relation $Q_{spec} \propto |S|$.

A more appropriate interpretation is:

$$Q_{spec} = f(K_c, S_t, D, I_a, T), \tag{13}$$

where $K_c$ – relevant knowledge coverage; $S_t$ – structural organization; $D$ – determinacy; $I_a$ – agent interpretability; $T$ – traceability.

### *C. Native and Foreign Specification Behavior*

The key experimental modification was to separate:

- native specification — generated within the target agent environment;
- foreign specification — generated within another agent environment.

The complete set of evaluated native and cross-agent specification combinations is summarized in Table III. Configurations that were not executed are marked as not evaluated.

TABLE II. TUNED SINGLE-AGENT MIGRATION PERFORMANCE

| Agent/environment | Token F1 | Exact match, % | SQL validity, % | AST exact, % | AST similarity | Runnable, % |
|---|---|---|---|---|---|---|
| Amazon Kiro | 0.56 | 12.04 | 57.8 | 12.98 | 0.40 | 14.76 |
| Spec Kit + Copilot | 0.73 | 12.76 | 73.03 | 12.65 | 0.55 | 17.48 |
| Claude Code | 0.73 | 12.87 | 56.21 | 12.49 | 0.41 | 4.11 |
| Antigravity + Gemini | 0.70 | 11.99 | 60.16 | 18.68 | 0.67 | 28.75 |
| Cursor | 0.69 | 12.10 | 65.98 | 16.11 | 0.44 | 7.49 |

TABLE III. CROSS-AGENT SPECIFICATION TRANSFER PERFORMANCE

| Implementation Agent | Specification Used | Token F1 | Exact Match, % | SQL Syntax Validity, % | AST Exact Match, % | AST Mean Similarity | Runnable, % |
|---|---|---|---|---|---|---|---|
| Kiro | Kiro | 0.365 | 0.33 | 33.74 | 0.33 | 0.2073 | 0.72 |
| Kiro | Gemini | 0.71 | 13.15 | 43.51 | 20.03 | 0.31 | 2.22 |
| Kiro | Copilot | — | — | — | — | — | — |
| Gemini | Kiro | 0.68 | 0.56 | 0 | 0 | 0.08 | 0.17 |
| Gemini | Gemini | 0.73 | 13.32 | 45.54 | 12.98 | 0.69 | 2.22 |
| Gemini | Copilot | — | — | — | — | — | — |
| Copilot | Kiro | 0.69 | 12.4 | 43.4 | 17.39 | 0.55 | 2.38 |
| Copilot | Gemini | 0.665 | 12.8 | 63 | **19.9** | **0.514** | Not implemented |
| Copilot | Copilot | 0.75 | 13.15 | 43.51 | **12.76** | **0.257** | 0 |

TABLE IV. REPLICATED GEMINI–KIRO TRANSFER RESULT

| Implementation agent | Specification origin | Token F1 | Exact match, % | SQL validity, % | AST exact, % | AST similarity | Runnable, % |
|---|---|---|---|---|---|---|---|
| Gemini | Kiro | **0.035** | **0.39** | **2.33** | **0.39** | **0.015** | **1.22** |

The results (Table III) show that the effect of specification transfer is asymmetric and agent-dependent. The most pronounced degradation occurred for Gemini using a Kiro-origin specification, whereas other foreign-specification combinations retained or improved individual metrics.

The present study addresses this gap through measured native- and foreign-specification experiments in an Oracle-to-PostgreSQL migration setting. One of the most important observations occurred when Gemini implemented from a Kiro specification.

The first cross-agent run showed substantial degradation. Because this result was anomalous, the Gemini–Kiro configuration was repeated independently; the replicated measurements are reported in Table IV. The replicated run confirmed the same qualitative failure pattern, with Token F1 of 0.035, SQL Syntax Validity of 2.33%, and AST Mean Similarity of 0.015.

However, the same effect was not universal across all agents. Copilot behaved substantially differently, in some conditions remaining relatively stable and in others benefiting from a foreign specification.

This supports the directional model:

$$C_{Kiro \to Gemini} \neq C_{Gemini \to Kiro} \tag{14}$$

and more generally:

$$C_{i \to j} \neq C_{j \to i}. \tag{15}$$

### *D. Specification Adaptation Strategies*

**Rewrite.** The implementation agent was asked to rewrite the foreign specification into its preferred representation before code generation.

For Gemini consuming Kiro specifications, rewriting substantially improved several metrics relative to the failed direct foreign-spec condition.

TABLE V. FOREIGN-SPECIFICATION REWRITING RESULTS

| Agent | Origin | Token F1 | Exact match, % | SQL validity, % | AST exact, % | AST similarity | Runnable, % |
|---|---|---|---|---|---|---|---|
| Gemini | Kiro | 0.68 | 13.15 | 43.62 | 13.10 | 0.22 | 0 |
| Copilot | Kiro | 0.60 | 0 | 14.09 | 0 | 0.034 | 0 |

The result is particularly informative because the knowledge source remained Kiro's specification, while its representation was transformed before implementation.

This suggests that part of cross-agent degradation may arise from representation compatibility rather than purely from missing knowledge.

**Specification Compression.** A second strategy reduced specification length through summarization.

TABLE VI. FOREIGN-SPECIFICATION COMPRESSION RESULTS

| Implementation Agent | Specification Origin | Token F1 | Exact Match, % | SQL Syntax Validity, % | AST Exact Match, % | AST Mean Similarity | Runnable, % |
|---|---|---|---|---|---|---|---|
| Copilot | Kiro | 0.72 | 12.71 | 43.90 | 19.48 | 0.24 | 0.22 |

The experiment did not provide evidence that simple length reduction universally solves specification incompatibility.

This is consistent with the earlier observation that length itself is not an adequate proxy for operational specification quality.

**Retrieval-Augmented Specification Ingestion.** In the third strategy, agents were prevented from directly reading the complete specification and instead received retrieval tools for selective specification access.

The RAG-based strategy did not outperform all alternative ingestion approaches across every metric. However, it was the only common strategy represented on the per-agent Pareto frontiers for both Gemini and Copilot, indicating the most consistent cross-agent trade-off among the evaluated strategies.

The reported results are shown in Table VII.

TABLE VII. RETRIEVAL-AUGMENTED SPECIFICATION INGESTION RESULTS

| Agent | Spec origin | Token F1 | Exact match, % | SQL validity, % | AST exact, % | AST similarity | Runnable, % |
|---|---|---|---|---|---|---|---|
| Gemini | Kiro | 0.57 | 13.40 | 13.82 | 20.20 | 0.32 | 1.80 |
| Copilot | Kiro | 0.69 | 12.76 | 43.51 | 12.99 | 0.285 | 0.28 |

RAG therefore did not simply dominate every other strategy on every metric. Its significance is different. To compare the observed strategies across agents, SQL Syntax Validity and AST Mean Similarity were considered jointly. These metrics represent complementary properties of the generated migration output: syntactic validity and structural similarity to the expected PostgreSQL implementation. Fig. 4 summarizes their behavior across increasing levels of specification foreignness and shows the corresponding global Pareto frontier.

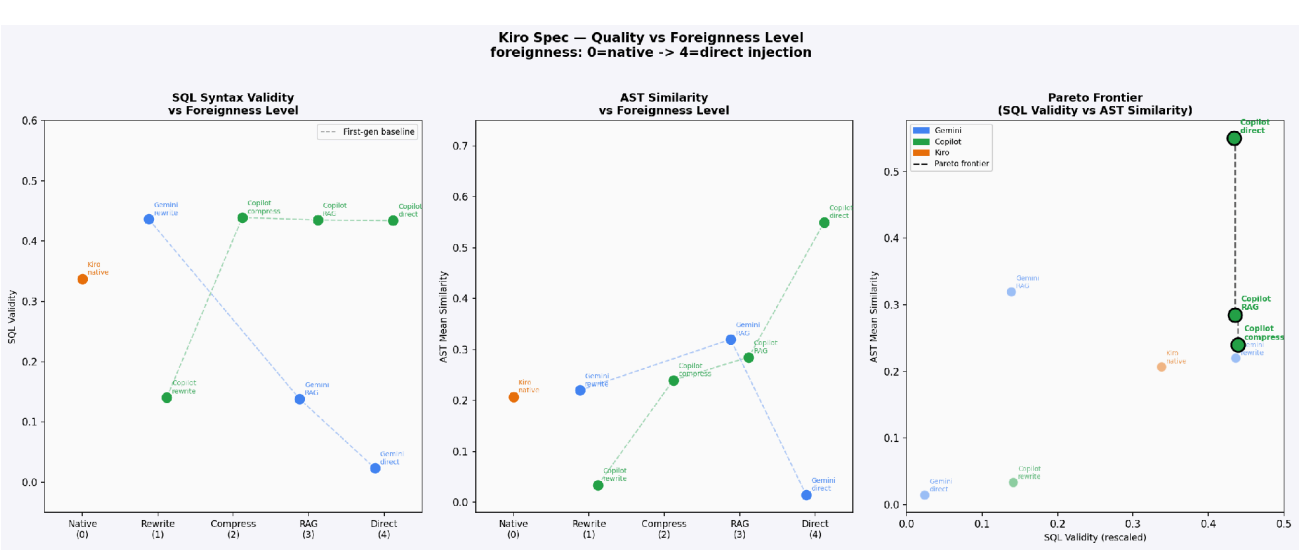

Fig. 4. Effect of specification foreignness and ingestion strategy on SQL syntax validity, AST similarity, and the global Pareto frontier.

The figure highlights different responses of Gemini and Copilot to increasing specification foreignness. Gemini exhibits substantial degradation under direct foreign-specification ingestion, whereas Copilot remains comparatively stable in several tested conditions. As a result, the global Pareto frontier is dominated by Copilot configurations. This difference indicates that cross-agent specification effects are agent-dependent rather than uniform across development environments.

Because the global frontier can be dominated by the generally stronger configurations of one agent, a per-agent comparison provides a more informative view of strategy robustness. Fig. 5 therefore evaluates the Pareto frontier separately for Gemini and Copilot when both operate on a Kiro-origin specification.

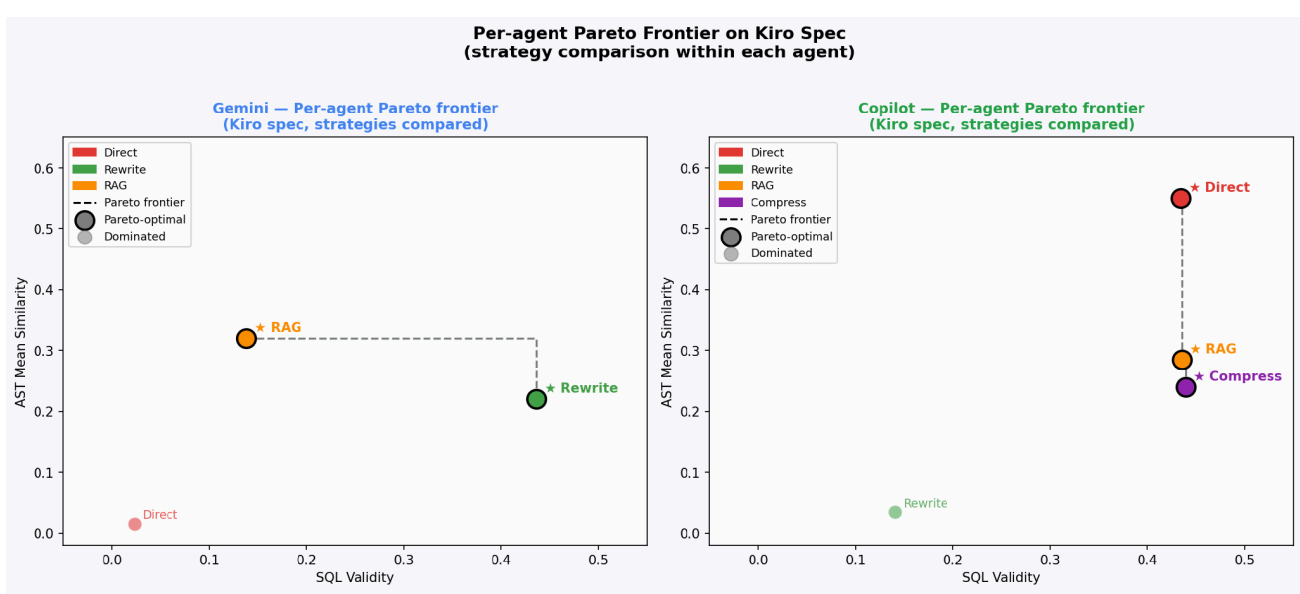

Fig. 5. Per-agent Pareto frontiers of specification-ingestion strategies for Gemini and Copilot using a Kiro-origin specification.

The per-agent analysis shows that the optimal strategy differs between implementation agents. However, RAG is the only evaluated ingestion strategy represented on the Pareto frontier of both agents. This does not establish RAG as universally optimal, but it identifies retrieval-based specification access as the most robust common strategy among the tested alternatives when cross-agent compatibility is unknown.

### *E. Proposed Specification Portability Model*

Based on the empirical results, we propose that SDD systems distinguish at least four properties of a specification:

$$S = \langle K, R, D, G \rangle, \tag{16}$$

where $K$ – knowledge contained by the specification; $R$ – representation and organization; $D$ – dependency structure; $G$ – granularity.

The effective specification received by an implementation agent is then:

$$S' = T(S, I_s, A_i), \tag{17}$$

where $T$ represents the ingestion/adaptation mechanism.

For direct ingestion: $T_{direct}(S) = S$.

For rewriting: $T_{rewrite}(S, A_i) = SA_i$.

For compression: $T_{compress}(S) = \hat{S}$.

For RAG: $T_{RAG}(S, q) = \{b_k \in S | relevance(b_k, q) > \theta\}$.

Thus implementation becomes:

$$Code_T = G(A_i, T(S, I_s, A_i), C).$$

This model distinguishes an important point: *"Specification information content and operational effectiveness are not identical"*.

Two specifications may express approximately the same functional intention but trigger substantially different downstream behavior in heterogeneous agents.

## F. Specification Portability as Knowledge Transition

The results can be connected to a broader SDD principle. A conventional view assumes: “Knowledge→Specification→Code”.

However, multi-agent SDD actually involves:

$$K \to S_{A_1} \to A_2 \to I_{A_2}.$$

At every transition, some portion of operational knowledge may become:

- preserved;
- reformulated;
- omitted;
- duplicated;
- ambiguously interpreted;
- given different priority.

We therefore distinguish:

Knowledge preservation: $K(S_{A_i}) \approx K(S_{A_j})$.

Operational interpretation preservation:

$$G(A_j, S_{A_i}) \approx G(A_j, S_{A_j}).$$

These distinctions explain why preserving specification content does not necessarily preserve implementation behavior across heterogeneous agents. The experimental results indicate that the same specification may be interpreted differently depending on the implementation agent and ingestion strategy. Therefore, specification portability should be evaluated not only in terms of information preservation, but also through the quality of the resulting implementation. This perspective provides the basis for interpreting the cross-agent results discussed in the following section.

# VIII. Discussion

The experiments show that specification quality cannot be explained by specification size alone. Kiro produced the largest specification, while Gemini generated a much shorter one, yet the larger specification did not consistently produce better code. More importantly, cross-agent transfer produced strongly agent-dependent results. Gemini showed substantial degradation when directly using a Kiro-origin specification, and this behavior was reproduced in a second run.

The adaptation experiments indicate that representation also matters. Rewriting the foreign specification substantially improved Gemini’s results, while the same strategy was less effective for Copilot. Compression did not provide a universal improvement. RAG also did not dominate all metrics, but it was the only common strategy appearing on the per-agent Pareto frontiers for both Gemini and Copilot.

Therefore, the results do not support a general claim that foreign specifications are inherently worse. Rather, they suggest that specification effectiveness depends on the interaction between its representation, its origin, and the implementation agent.

# IX. Implications For Multi-Agent Software Engineering

The findings have direct implications for multi-agent development workflows. A specification successfully used by one agent should not automatically be treated as equally suitable for another agent. Agent substitution may therefore require specification adaptation or an alternative ingestion strategy.

The results also suggest that multi-agent SDD systems should distinguish between the semantic content of a specification and the way this content is presented to a particular agent. Instead of maintaining several completely independent specifications, a common specification could be preserved while different agents receive adapted or selectively retrieved views of it.

Among the evaluated strategies, retrieval-based access appears to be the most robust common option when compatibility between agents is unknown, although the current experiments are not sufficient to claim that RAG is universally optimal.

# X. Connection With A Formal Specification Model

The cross-agent results also support the use of a structured rather than purely linear specification representation. A specification can be considered as a set of semantic blocks containing requirements, rules, dependencies, decisions, and unresolved questions. An implementation agent does not necessarily need the complete specification for every task; it may instead receive only the blocks relevant to the current implementation object and their dependencies.

This interpretation provides a natural extension of the current RAG experiment. Instead of retrieving fragments only through textual or semantic similarity, future implementations could retrieve specification blocks according to explicit dependency relations. Such an approach could reduce unnecessary context while preserving the information required for implementation.

Thus, the formal specification model and the cross-agent experiments lead to a common architectural direction: maintain a structured specification as the stable source of knowledge and provide agent-specific views of this specification during implementation. This could reduce dependence on the writing style or context organization of a particular LLM agent.

# XI. Conclusion

This study investigated whether specifications used in LLM-based specification-driven development can be reliably transferred between heterogeneous development agents. The initial Oracle-to-PostgreSQL migration experiment confirmed the practical feasibility of a specification-first approach: 623 of 1,006 PL/SQL files were successfully regenerated, and 380 of the regenerated scripts executed successfully in PostgreSQL 16.

The cross-agent experiments showed that specification effectiveness is agent-dependent. Specification size alone did not explain implementation quality, and transferring a specification between agents could substantially change the result. The strongest observed case was the degradation of Gemini when directly consuming a Kiro-origin specification, which was reproduced in a repeated run.

The evaluated adaptation strategies produced different effects across agents. Rewriting substantially improved Gemini’s results in the tested configuration, while compression did not provide a universal benefit. RAG did not outperform all alternatives on every metric, but it was the only common ingestion strategy represented on the per-agent Pareto frontiers for both Gemini and Copilot.

Overall, the results suggest that specifications in heterogeneous SDD workflows should not automatically be treated as agent-neutral artifacts. Their effectiveness depends not only on the information they contain, but also on how that information is structured, transferred, and interpreted by the implementation agent. This motivates further research on specification portability, structured specification representations, agent-specific adaptation, and dependency-aware retrieval for multi-agent software engineering.